\documentclass[aps,prl,reprint,amsmath,amssymb,floatfix,superscriptaddress]{revtex4-2}
\usepackage[justification=Justified,singlelinecheck=false]{caption}
\usepackage{bbm}
\usepackage{graphicx}
\usepackage{physics}
\usepackage{upgreek}
\usepackage{siunitx}
\usepackage{hyperref}
\renewcommand{\selectlanguage}[1]{}
\usepackage{verbatim}

\newcommand{\startSequations}{
  \setcounter{equation}{0}
  \renewcommand{\theequation}{S\arabic{equation}}
}

\newcommand{\startSfigures}{
  \setcounter{figure}{0}
  \renewcommand{\thefigure}{S\arabic{figure}}
  \setcounter{table}{0}
  \renewcommand{\thetable}{S\arabic{table}}
}

\newcommand{\affA}{ETH Zürich – PSI Quantum Computing Hub, Paul Scherrer Institut (PSI), 5232 Villigen, Switzerland}
\newcommand{\affB}{Institute for Quantum Electronics, ETH Zürich, Otto-Stern-Weg 1, 8093 Zürich, Switzerland}
\newcommand{\affC}{Quantum Center, ETH Zurich, CH-8093, Switzerland}
\newcommand{\affD}{Department of Physics, University of Trieste, 34127 Trieste, Italy}

\newcommand{\equalcontrib}{\thanks{These authors contributed equally to this work.}}

\begin{document}

\title{State-dependent Gaussian gate set using an optical tweezer for trapped ions}

\author{Philip Leindecker}
\email[]{pleindecker@ethz.ch}
\equalcontrib
\affiliation{\affA}\affiliation{\affB}

\author{Luka Milanovic}
\equalcontrib
\affiliation{\affA}\affiliation{\affB}

\author{Tanja Behrle}
\affiliation{\affA}

\author{Edgar Brucke}
\affiliation{\affA}\affiliation{\affB}

\author{Matteo Marinelli}
\affiliation{\affA}\affiliation{\affB}\affiliation{\affD}

\author{Julian Schmidt}
\affiliation{\affA}

\author{Jonathan Home}
\affiliation{\affB}\affiliation{\affC}

\author{Cornelius Hempel}
\affiliation{\affA}\affiliation{\affB}\affiliation{\affC}

\date{\today}

\begin{abstract}
We demonstrate a state-dependent Gaussian gate set on the motional modes of trapped $^{40}$Ca$^+$ ions, realized with an optical tweezer.
Dynamic control of the tweezer intensity and position enables local displacement, squeezing, phase-space rotation, and beamsplitter operations, constituting a complete gate set.
By varying the tweezer position relative to the ion, we show how the strength of each operation is set by the corresponding spatial derivative of the local optical potential.
We further demonstrate the inherent dependence of each operation on the ion's internal state and use coherent spin-motion coupling provided by the tweezer to create a motional cat state.
Our work establishes optical tweezers as a unified and local resource for continuous-variable quantum control in trapped ion systems.
\end{abstract}

\maketitle

Continuous-variable quantum information processing based on bosonic modes offers a complementary paradigm to discrete-variable approaches by encoding information in the infinite-dimensional Hilbert space of quantum harmonic oscillators~\cite{braunstein_quantum_2005}.
Bosonic modes have been explored in different architectures including photonic circuits~\cite{enomoto_programmable_2021, arrazola_quantum_2021}, superconducting microwave cavities~\cite{eickbusch_fast_2022}, and trapped ions~\cite{fluhmann_encoding_2019, de_neeve_error_2022, fontbote-schmidt_error_2026}.
In this framework, a Gaussian gate set, comprising displacement, squeezing, phase-space rotation, and beamsplitter operations, together with at least one non-Gaussian element, is universal~\cite{lloyd_quantum_1999}.
Implementing these operations on individual bosonic modes is a prerequisite for realizing bosonic quantum simulators~\mbox{\cite{lau_proposal_2012, tamura_quantum_2020, debnath_observation_2018, katz_programmable_2023, somhorst_quantum_2023}}, which can be extended to hybrid oscillator-qubit architectures~\cite{andersen_hybrid_2015, sutherland_universal_2021, gallagher_quadratic_2025, liu_hybrid_2026} by including spin degrees of freedom.

In atomic platforms, such as neutral atoms and trapped ions, tightly focused laser beams, also known as optical tweezers, can provide unified control over bosonic modes via a light-induced potential.
For neutral atoms, tweezers generate the dominant confinement of the atoms~\cite{grimm_optical_2000} and have been used to realize large-scale, dynamically reconfigurable arrays of thousands of individually trapped atoms~\cite{manetsch_tweezer_2025}. 
Additionally, recent proposals for neutral atoms suggest modulating the position and intensity of the confining tweezer potential to realize continuous-variable operations~\cite{grochowski_quantum_2025, hwang_hybrid_2025}.
Optical tweezers have also been proposed in the context of ion chains, where they can be used to modify the normal mode structure to mitigate mode crowding~\mbox{\cite{olsacher_scalable_2020, shen_scalable_2020, teoh_manipulating_2021, schwerdt_scalable_2024}} or to implement new types of gates~\cite{mazzanti_trapped_2021, mazzanti_trapped_2023}, of which first demonstrations were recently reported~\cite{schwerdt_optical_2026, cui_transverse_2025, faorlin_entangling_2026}.
Nonharmonic terms in the tweezer potential due to its tight focus can enable non-Gaussian control~\cite{grochowski_quantum_2025}, but also limit the achievable size of motional states~\cite{drechsler_state-dependent_2020}.
Since the tweezer operations are inherently state-dependent, the spin can also provide the necessary nonlinearity to render them non-Gaussian~\cite{lau_universal_2016}.

Individual Gaussian operations have been implemented using several techniques with trapped ions.
One approach is to use parametric modulation of the electric trapping potential~\cite{burd_quantum_2019, metzner_two-mode_2024}, which typically employs a global drive and lacks individual ion addressing capabilities.
Alternatively, such operations have been implemented by optically driving motional sideband transitions~\cite{millican_engineering_2025, bazavan_squeezing_2026}.
While this allows for local control and state dependence, the trade-offs can include constraints on the beam geometry~\cite{leibfried_quantum_2003} and off-resonant driving of the carrier transition.
The latter is especially relevant for the beamsplitter operating at the frequency difference between modes~\cite{gan_hybrid_2020, jeon_two-mode_2025}, which is a central primitive for direct implementation of energy-preserving logical gates on GKP-encoded qubits~\cite{rojkov_two-qubit_2024}.
Instead, using the potential induced by far off-resonant dipole forces as a unified resource for the individual operations offers an additional toolkit which can be applied locally using tight focusing.

In this Letter, we use a single optical tweezer to demonstrate an internal-state dependent Gaussian gate set on the motional modes of trapped $^{40}$Ca$^+$ ions.
The relevant confinement of the ions is provided by electric potentials of the ion trap, with a relatively weak modification of the trapping potential by the tweezer.
By arbitrarily positioning the tweezer relative to a single ion, we can realize different interactions which depend on the gradient and curvature of the tweezer intensity.
In particular, we show the dependence of the displacement operation on the local gradient of the tweezer potential and the squeezing operation on its local curvature.
We resonantly drive those operations by modulating the intensity of the tweezer light at their respective frequencies and verify the prepared motional states by direct characteristic function readout~\cite{fluhmann_direct_2020}.
The static curvature of the tweezer potential leads to a shift of the motional mode frequency, which also acts as a rotation operation in phase space.
Each operation is inherently dependent on the internal state, which we characterize and subsequently exploit to prepare a coherent spin-motion cat state, illustrating the capability of non-Gaussian operations via the spin.
We complete the state-dependent Gaussian gate set by implementing a state-dependent beamsplitter on two motional modes in a two-ion crystal via modulating the intensity of the tweezer light at the frequency difference of the two motional modes.

The interaction of a two-level system with an off-resonant optical tweezer, which couples the states $\ket{i} \in \{ \ket{\uparrow}, \ket{\downarrow} \}$ to auxiliary states through ac-Stark shifts, is described by the Hamiltonian $\hat{H}_{\rm int} = \sum_i\Delta E_i \ket{i}\bra{i}$ with $\Delta E_i = \hbar\alpha_i(\lambda, \theta)I(x, y) / (2c\epsilon_0)$~\cite{grimm_optical_2000}.
The polarizability $\alpha_i(\lambda, \theta)$ depends on the wavelength $\lambda$ and the polarization angle $\theta$ of the tweezer beam with respect to the magnetic field, and $I(x, y)$ is a Gaussian intensity profile of the tweezer beam with transverse coordinates $(x, y)$.
Expanding $\Delta E_i$ to second order with respect to the spatial coordinates around the center of the trapping potential $(x_0,y_0=0)$ gives
\begin{align}
    \Delta E_i 
    \approx \frac{\alpha_i}{2c\epsilon_0} 
    \bigg[ 
        I(x_0) + 
        \left. \partial_x I(x) \right|_{x_0} \hat{x} +
        \frac{1}{2} \left. \partial_x^2 I(x) \right|_{x_0} \hat{x}^2
    \bigg].
\end{align}
The zero-order term $\propto I(x_0)$ describes the ac-Stark shift of the energy levels $i$ by $h \Delta_{\rm ac, i}$, while the first-order term $\propto \partial_x I(x)|_{x_0}$ induces a force and the second-order term $\propto \partial_x^2 I(x)|_{x_0}$ corresponds to the curvature of the state-dependent potential $\sim \omega_{{\rm tw}, i}^2$.
The individual state-dependent Gaussian operations can be derived from those terms and all inherit the general form $\hat{U}_O = \sum_i \hat{O}(p_i)\ket{i}\bra{i}$, where the parameter $p_i$ sets the operation strength for each internal state $\ket{i}$.
We refer the reader to the Supplemental Material (SM)~\cite{supplemental} \nocite{atomphys, kiruga_portal_2025, james_quantum_1998, leindecker_direct_2026, fluhmann_encoding_2019_phd_thesis} for a detailed derivation of each operation.

\begin{figure}[t!]
    \centering
    \includegraphics[]{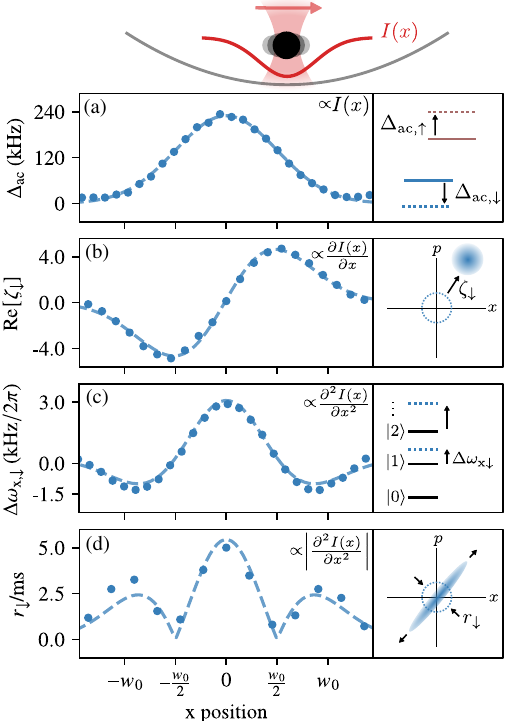}
    \caption{\textbf{Position-dependent tweezer operations.}
    We measure the tweezer-induced (a) differential ac-Stark shift $\Delta_{\rm ac}$, (b) displacement $\zeta_{\downarrow}$, (c) axial motional mode shift $\Delta \omega_{x, \downarrow}$, and (d) squeezing strength $r_{\downarrow}/ \rm ms$ along the $x$-direction aligned with the axial motional mode and the ion at $x=0$, illustrating the different derivatives of the tweezer potential from top to bottom.
    The ac-Stark shift in (a) is measured by spectroscopically probing the $\ket{\downarrow} \leftrightarrow \ket{\uparrow}$ transition and is directly proportional to the intensity profile $I(x)$ of the tweezer beam.
    The displacement and squeezing parameters in (b) and (d) are determined via characteristic function readout~\cite{fluhmann_direct_2020} after modulating the intensity of the tweezer at $\omega_{x}$ for $\SI{20}{\micro\second}$ and $\omega_{\rm sq}$ for $\SI{200 - 600}{\micro\second}$, respectively. 
    The motional mode shift $\Delta \omega_{x, \downarrow}$ in (c) is measured via resonant excitation of the ion motion and subsequent application of a RSB $\pi$-pulse of fixed duration followed by state detection, allowing for a temperature-sensitive measurement.
    The dashed lines in each subplot are the calculated values for the terms up to second order in $\hat{H}_{\rm int}$, for a Gaussian beam with a waist of $w_0 = \SI{1.38(1)}{\micro\meter}$, extracted from (a), normalized to the measured data (dots), showing good agreement between theory and experiment for the different operations given by the tweezer potential.
    Error bars are smaller than the markers.}
    \label{fig:position_dependence}
\end{figure}

We first demonstrate the single-mode operations using the axial motional mode with frequency $\omega_x$ of a single $^{40}$Ca$^+$ ion confined in a linear Paul trap with secular frequencies $(\omega_x,\omega_y,\omega_z) = 2\pi\times \SI{(1.19,\,2.02,\,2.36)}{\mega\hertz}$.
Our optical tweezer with a waist of $w_0 \approx \SI{1.38(1)}{\micro\meter}$ is tuned to a wavelength of $\approx \SI{733}{\nano\meter}$ and mainly off-resonantly couples to the dipole-allowed transitions near $\SI{393}{\nano\meter}$ and $\SI{397}{\nano\meter}$ (for $\ket{\downarrow} \leftrightarrow \,\{ ^4P_{3/2}, \, ^4P_{1/2} \}$) and $\SI{854}{\nano\meter}$ (for $\ket{\uparrow} \leftrightarrow \,^4P_{3/2}$).
The polarization of the tweezer is linear along the $y$-axis to avoid undesired mixing between neighboring Zeeman levels, which is especially relevant in the high-power tweezer regime~\cite{leindecker_anti_crossing}.
Two crossed acousto-optical deflectors (AODs) position the tweezer with respect to the ion in the $x$\nobreakdash-$y$~plane.
A single-pass acousto-optical modulator (AOM) located before the AODs is used for amplitude modulation of the tweezer light.
For each measurement, the ions are first cooled close to the motional ground state by a combination of dark resonance and resolved sideband cooling with the tweezer light off, followed by state preparation in either the $\ket{\downarrow} = \ket{4S_{1/2},\,m_j=+1/2}$ or $\ket{\uparrow} = \ket{3D_{5/2},\,m_j=+3/2}$ state, using a narrow-linewidth quadrupole laser at $\approx \SI{729}{\nano\meter}$.
Next, we turn on the tweezer for a duration $t$ at the target location with up to $\approx \SI{100}{\milli\watt}$ at the ion and optional intensity modulation.
State discrimination is performed at the end of the sequence via state-dependent fluorescence~\cite{leibfried_quantum_2003} on the dipole transition at $\SI{397}{\nano\meter}$.
More details can be found in the SM~\cite{supplemental}.

In a first set of measurements, we demonstrate how the strengths of the operations associated with the terms up to second order in $\hat{H}_{\rm int}$ depend on the position of the tweezer $x_{\rm tw}$ relative to the ion.
Specifically, we map out the spatial dependence of the ac-Stark shift, displacement and squeezing operations by scanning the tweezer position along the axial direction of the trap, while the ion stays located at $x=0$ for all measurements.
We measure the differential ac-Stark shift $\Delta_{\rm ac}$, see Fig.~\ref{fig:position_dependence}~(a), by spectroscopically probing the $\ket{\downarrow} \leftrightarrow \ket{\uparrow}$ transition using the resonant $\SI{729}{\nano\meter}$ beam which propagates along the $x$-axis, while simultaneously shining tweezer light with $< \SI{1}{\milli\watt}$ onto the ion.
The resulting spatial profile confirms the Gaussian envelope of the tweezer beam.

The displacement operation is derived from the force term in $\hat{H}_{\rm int}$ and is given by $\hat{U}_D=\sum_i \hat{D}(\zeta_i) \ket{i}\bra{i}$, with $\hat{D}(\zeta_i) = \exp \left ( \zeta_i \hat{a}^\dagger - \zeta_i^* \hat{a} \right)$ of amplitude $\zeta_i$, where $\hat{a}^\dagger$, $\hat{a}$ are the creation and annihilation operators.
By modulating the intensity $I(x,y)$ at the axial trap frequency $\omega_x$ of the ion, we can resonantly drive this operation.
For every tweezer position along $x$, we turn on the displacement operation for a fixed time of $\SI{20}{\micro\second}$, followed by state tomography via characteristic function readout using the axial $\SI{729}{\nano\meter}$ beam~\cite{fluhmann_direct_2020}.
Fitting the imaginary part of the characteristic function to an ideal displaced state $\ket{\zeta_\downarrow} = \hat{D}(\zeta_\downarrow)\ket{0}$ allows us to determine the strength and the sign of the displacement, thereby revealing the gradient of the tweezer light field, as shown in Fig.~\ref{fig:position_dependence}~(b).
At the positions of maximum gradient, at $x_{\rm tw} = \pm w_0/2$, we observe maximum displacements with opposite signs on either side of the focus, and a vanishing displacement at the beam center, $x_{\rm tw}=0$.

The second-order term in $\hat{H}_{\rm int}$ causes a motional mode shift, which, for the axial mode $\omega_x$ and state $\ket{\downarrow}$, is given by $\Delta \omega_{x, \downarrow} = \sqrt{\omega_x^2 + \omega_{\rm tw, \downarrow}^2} - \omega_{x}$.
We measure the shift by turning on the tweezer and simultaneously applying a weak electric field modulated at frequency $\omega_{\rm tickle}$ to a microwave antenna line integrated in the ion trap for a duration of $\SI{1.5}{\milli\second}$~\cite{vasquez_state-dependent_2024}.
After this pulse, we probe the red-sideband (RSB) transition for a duration of $\SI{32}{\micro\second}$ using the axial $\SI{729}{\nano\meter}$ beam as a function of  $\omega_{\rm tickle}$ and the position of the tweezer to detect motional excitation out of the initially prepared ground state.
The extracted motional mode shift $\Delta \omega_{x, \downarrow}$ as a function of tweezer position is shown in Fig.~\ref{fig:position_dependence}~(c).
The modulation at $\omega_{\rm tickle}$ is kept sufficiently low to keep the extent of the motional wavefunction close to that of the ground state, $\approx \SI{10}{\nano\meter}$, which is much smaller than the waist of the tweezer, ensuring that higher-order terms of the optical potential are negligible.
We determine a maximal motional mode shift in the center of the beam as well as a negative shift at $x_{\rm tw} = \pm \sqrt{3}w_0/2$.
This shift corresponds to a rotation operation in phase space $\hat{U}_R=\sum_i \hat{R}(\varphi_i) \ket{i}\bra{i}$, with $\hat{R}(\varphi_i) = \exp \left ( -i\varphi_i\hat{a}^\dagger\hat{a}\right )$, by an angle $\varphi_i = \Delta\omega_{i} t$, where $t$ is the duration of the tweezer pulse.

The second-order term also enables a squeezing operation, given by $\hat{U}_S=\sum_i \hat{S}(\xi_i) \ket{i}\bra{i}$ with $\xi_i = r_i e^{i\phi}$, where $r_i$ describes the degree of squeezing along a phase space direction at angle $\phi$.
Similarly to the displacement operation, modulating the tweezer intensity at frequency $\omega_{\rm sq} = 2 \left( \omega + \Delta \omega_{i}/2 \right )$ resonantly drives this operation.
The offset $\Delta \omega_{i}/2$ accounts for the time-averaged motional mode shift present during squeezing.
We choose the squeezing duration between $\SI{200}{\micro\second}$ and $\SI{600}{\micro\second}$ to compensate for weaker squeezing away from the center of the beam and normalize the measured squeezing parameter to determine the squeezing strength, which is plotted as a function of the tweezer position in Fig.~\ref{fig:position_dependence}~(d).
The parameter $r$ is determined by fitting the real part of the characteristic function to an ideal squeezed state $\ket{\xi_\downarrow} = \hat{S}(\xi_\downarrow)\ket{0}$.
We find the squeezing to be maximum at $x_{\rm tw} = 0$, and vanishing around $x_{\rm tw} = \pm w_0/2$.
In contrast to the motional mode shift, which is proportional to the curvature of the optical potential generated by the tweezer, the measured squeezing strength is proportional to the absolute value of the curvature.
All four measured effects are in good agreement with the theoretical description by expanding $\hat{H}_{\rm int}$ to second order as indicated by the dashed lines in each subfigure of Fig.~\ref{fig:position_dependence}.

\begin{figure*}
    \centering
    \includegraphics[]{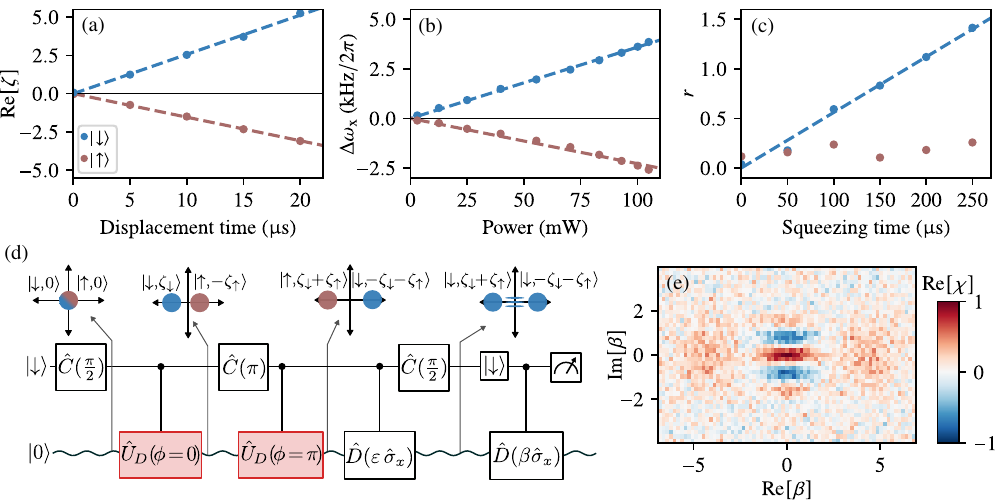}
    \caption{\textbf{State-dependent tweezer operations}. 
    (a) Displacement as a function of modulation duration for the states $\ket{\downarrow}$ (blue) and $\ket{\uparrow}$ (brown) at the position of maximum gradient, $x_{\rm tw} = w_0/2$.
    The two states are displaced in opposite directions in phase space due to opposite signs of the gradients of the light field.
    (b) Axial motional mode shift at the center of the tweezer as a function of tweezer power.
    The potential generated by the tweezer is positive for the $\ket{\downarrow}$ state and negative for the $\ket{\uparrow}$ state.
    (c) Squeezing as a function of squeezing modulation time. 
    This operation is frequency selective, in contrast to the displacement operation, which allows squeezing of one state (here $\ket{\downarrow}$) while leaving the other state ($\ket{\uparrow}$) unperturbed.
    Error bars in (a)–(c) are smaller than the markers.
    (d) Experimental sequence and phase space illustration for reading out a motional cat state using the spin-dependent displacement operation of the tweezer (marked in red) followed by two SDF pulses using the $\SI{729}{\nano\meter}$ axial beam to disentangle spin and motion and to measure the characteristic function, with a short spin reset in between for improved contrast.
    (e) Real part of the characteristic function of a motional cat state of size $2\zeta_{\rm cat} = 4.05(2)$, prepared and read out using the pulse sequence illustrated in (d).}
    \label{fig:state_dependence}
\end{figure*}

All presented tweezer operations are inherently internal-state dependent, given by the general form of the Hamiltonian $\hat{H}_{\rm int}$.
We next demonstrate this state dependence of each operation individually at the tweezer position that maximizes the corresponding term in $\hat{H}_{\rm int}$, using the same pulse sequence as in the position-dependence characterization.
To measure the operations for the $\ket{\uparrow}$ state, we apply a carrier $\pi$-pulse before and after the tweezer operation using the axial $\SI{729}{\nano\meter}$ beam.
For the displacement operation, we determine an amplitude of up to $\zeta_\downarrow = 5.2(1)$ for $\ket{\downarrow}$ and $\zeta_\uparrow = -3.09(8)$ for $\ket{\uparrow}$ at a tweezer modulation time of $\SI{20}{\micro\second}$, illustrated in Fig.~\ref{fig:state_dependence}~(a), confirming the opposite displacement direction in phase space for the two states, given by the different sign of the polarizabilities of the two states $\ket{i}$.
For the motional mode shift, we measure up to $\Delta \omega_{x, \downarrow} = 2\pi \times \SI{3.86(2)}{\kilo\hertz}$ and $\Delta \omega_{x, \uparrow} = -2\pi \times \SI{2.59(1)}{\kilo\hertz}$ for the two states, and illustrate the linear dependence on power in Fig.~\ref{fig:state_dependence}~(b).
In Fig.~\ref{fig:state_dependence}~(c), we tune the modulation frequency of the tweezer intensity to be optimal for squeezing the $\ket{\downarrow}$ state and reach up to $r_\downarrow = 1.41(4)$.
Due to the differential motional mode shift, the $\ket{\uparrow}$ state is effectively not squeezed at this modulation frequency.
Residual off-resonant squeezing can be minimized by increasing the motional mode shift.

The state dependence turns the Gaussian operations into a resource for coherent spin-motion control, which we demonstrate in the following by preparing a cat state.
Starting from $\ket{\downarrow, n=0}$, we apply the sequence $\hat{C}(\pi/2) \to \hat{U}_D(\phi=0) \to \hat{C}(\pi) \to \hat{U}_D(\phi=\pi)$ to prepare the cat state $\ket{\psi_{\rm cat, \downarrow\uparrow}} = \frac{1}{\sqrt{2}} \left( \ket{\downarrow, \zeta_\mathrm{cat}} + \ket{\uparrow, -\zeta_\mathrm{cat}} \right)$ with $\zeta_\mathrm{cat} = \zeta_\downarrow + \zeta_\uparrow$, as shown in Fig.~\ref{fig:state_dependence}~(d). 
Here $\hat{C}(\nu)$ denotes a $\hat{\sigma}_x$ rotation of the internal-state Bloch vector by an angle $\nu$, $\phi$ is the motional phase for $\hat{U}_D$, and the $\zeta_i$ are taken to be positive real numbers by convention.
The two displacement operations are applied for a duration of $\SI{16.8}{\micro\second}$ at a reduced power of $P\approx \SI{40}{\milli\watt}$.
The pulse $\hat{C}(\pi)$ acts as a spin-echo, ensuring equal magnitudes of displacement in phase space for the two spin states and canceling the ac-Stark shift accumulated during the displacements.
In order to read out the resulting state, we disentangle the spin from the motion via a short additional state-dependent force (SDF) pulse of duration $\SI{9}{\micro\second}$ using the $\SI{729}{\nano\meter}$ axial beam, leading to a negligible displacement $\varepsilon \ll \zeta_{\rm cat}$, followed by a final $\hat{C}(\pi/2)$ and state-preparation pulse of duration $\SI{20}{\micro\second}$~\cite{hastrup_measurement-free_2021}.
We read out the real part of the characteristic function, shown in Fig.~\ref{fig:state_dependence}~(e), and verify the expected cat state $\ket{\psi_{\rm cat, \downarrow}}$ with a fitted size of $2\zeta_\mathrm{cat} = 4.05(2)$.

\begin{figure}[ht]
    \centering
    \includegraphics[]{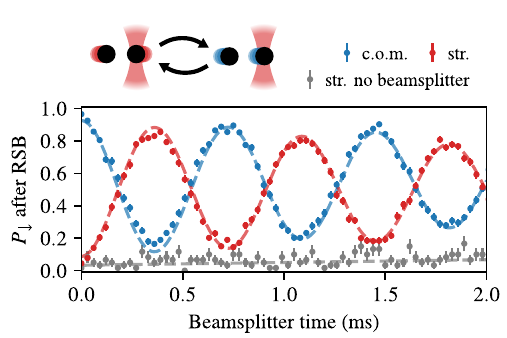}
    \caption{\textbf{Beamsplitter interaction in a two-ion crystal.} 
    Population exchange between the axial c.o.m\@.\ (blue) and str\@.\ (red) modes as a function of the beamsplitter interaction time. 
    The str\@.\ mode is initially prepared in Fock state $\ket{n=1}$, while the  c.o.m\@.\ mode is kept close to $\ket{n=0}$.
    The tweezer is centered on one ion and intensity modulated at the frequency difference between the two modes $\omega_{\rm str} - \omega_x$, driving a coherent swap of a single phonon between the two modes.
    We fit a time of $t_{\pi} = \SI{363(13)}{\micro\second}$ for a full population transfer between the two modes.
    The reduced initial contrast of the oscillations can be attributed to an imperfect RSB $\pi$-pulse for state preparation and readout, as indicated in gray, which was measured with a wait time applied instead of the beamsplitter pulse.
    The decay of the oscillations is consistent with the independently measured motional coherence.
    Error bars, where not visible, are smaller than the marker size.}
    \label{fig:beamsplitter}
\end{figure}

The size and fidelity of the cat state, as well as the achievable maximal values for $\zeta$ and $r$ for the individual operations, are limited by tweezer-induced errors.
In our setup, operating at the reduced tweezer power of $P \approx \SI{40}{\milli\watt}$ yields a spin coherence time of $T_2^s = \SI{98(2)}{\micro\second}$ at the steepest gradient, sufficient for the cat state preparation, and $T_2^s = \SI{153(3)}{\micro\second}$ in the center of the tweezer.
We find a motional coherence time, measured via the Fock state superposition $(\ket{0} + \ket{1})/\sqrt{2}$, at $\approx \SI{75}{\milli\watt}$ tweezer power of $T_2^m = \SI{9.0(4)}{\milli\second}$ at the center of the beam and $T_2^m = \SI{3.3(2)}{\milli\second}$ at its maximum gradient, while the heating rate remains unaffected.
The larger reduction of both coherence times at the steepest gradient indicates that beam-pointing fluctuations dominate over intensity fluctuations of the tweezer.
While the presented operations are theoretically reversible, these tweezer-induced errors limit their experimental reversibility.
Additionally, nonharmonic terms in the tweezer potential become relevant for large motional states, once the wavepacket size approaches the tweezer waist~\cite{drechsler_state-dependent_2020}.
For the characteristic function readout itself, effects of being outside the Lamb-Dicke regime pose the main limitation on the maximum size of motional states that can be reconstructed~\cite{simoni_non-linear_2025}.

Finally, in order to complete the universal gate set suited for multi-mode systems, we implement a beamsplitter interaction $\hat{B}(g_i) = \exp\left (\frac{1}{2} (g_i\hat{a}^{\dagger}\hat{b} - g_i^{*}\hat{a}\,\hat{b}^{\dagger} )\right )$ of strength $g_i$ on two modes, where $\hat{b}^\dagger$, $\hat{b}$ are the creation and annihilation operators of the second motional mode.
This operation is enabled by the same curvature term in $\hat{H}_{\rm int}$ as for the single-mode operations.
We drive the beamsplitter on a two-ion crystal between its axial center-of-mass (c.o.m\@.)\ mode $\omega_x$ and stretch (str\@.)\ mode $\omega_{\rm str} = 2\pi\times \SI{2.07}{\mega\hertz}$, with both modes cooled close to the motional ground state and at reduced radial frequencies $(\omega_y,\omega_z)=2\pi\times \SI{(1.87,\,2.25)}{\mega\hertz}$.
To observe Rabi oscillations of a single excitation between the two modes as a function of beamsplitter time, we first use the local ac-Stark shift of the tweezer to mask one of the ions (ion 1), while shelving the other ion (ion 2) into the auxiliary $\ket{3D_{5/2},\,m_j=+5/2}$ state using an additional $\SI{729}{\nano\meter}$ beam propagating at $40$ degrees to the trap axis.
Next, we prepare a Fock state $\ket{n=1}$ in the str\@.\ mode, while keeping the c.o.m\@.\ mode close to $\ket{n=0}$, by applying a carrier $\pi$-pulse and RSB $\pi$-pulse using the axial $\SI{729}{\nano\meter}$ beam acting on ion 1.
To drive the beamsplitter interaction, we shine the tweezer on ion 1 and modulate the tweezer intensity at $\omega_{\rm BS} = \omega_{\rm str} - \omega_x$ for a duration of $t_{\rm BS}$.
To read out the population in each motional mode, we apply a RSB $\pi$-pulse resonant with the respective transition.
The populations of the motional states as a function of beamsplitter interaction time are shown in Fig.~\ref{fig:beamsplitter}.
We determine $t_{\pi} = \SI{363(13)}{\micro\second}$, the time required for complete population transfer between the two modes, corresponding to a beamsplitter coupling strength of $g_\downarrow = 2\pi\times\SI{1.38(5)}{\kilo\hertz}$.
The initial contrast of the oscillations is limited by an imperfect RSB $\pi$-pulse (fidelity $\sim 93\%$), while the decay is consistent with the independently measured motional coherence time of the c.o.m\@.\ mode.

The demonstrated state-dependent gate set constitutes a new resource for bosonic quantum information processing with trapped ions, offering an alternative to existing approaches based on global electrical or optical drives.
Tweezer-based control can be straightforwardly extended to operations between arbitrary pairs of axial motional modes in long ion strings, to radial modes, and to two-mode squeezing.
The main limitations to operational fidelity, beam pointing fluctuations and intensity noise, could be mitigated with the use of integrated optics~\cite{vasquez_state-dependent_2024} or Laguerre-Gaussian beamshapes~\cite{mazzanti_trapped_2023}.
The strength of the demonstrated operations could be further increased by tighter focusing of the tweezer beam or by a more suitable choice of the tweezer wavelength to enhance the polarizability of selected internal states.
The demonstrated operations can aid in proposed sensing applications~\cite{drechsler_state-dependent_2020, bond_optimal_2025} as well as in two-qubit gates~\cite{shapira_robust_2023} or in splitting and transport in multi-ion crystals~\cite{sutherland_motional_2021}.

\begin{acknowledgments}
\textit{Acknowledgments}.--We thank Matteo Simoni, Alfredo Ricci-V\'{a}squez, Moritz Fontbot\'{e}-Schmidt, and Wojciech Adamczyk for helpful discussions. 
This work was supported by the ETH Zurich–PSI Quantum Computing Hub; the Intelligence Advanced Research Projects Activity (IARPA) and the Army Research Office under the Entangled Logical Qubits program through Cooperative Agreement No.~W911NF-23-2-0216.
\end{acknowledgments}

\bibliography{references}

\clearpage

\onecolumngrid

\begin{center}
  \vspace*{0.5\baselineskip}
  {\large\bfseries
   Supplemental Material for:\\[0.4em]
   State-dependent Gaussian gate set using an optical tweezer for trapped ions\par}
\end{center}

\vspace{1em}

\twocolumngrid

\startSequations
\startSfigures

\section{Tweezer-ion interaction}\label{sec:tweezer_ion_interaction}

\subsection{Hamiltonian}

The interaction of a two-level system with an optical field of intensity $I(x, y)$, off-resonantly coupling the states $i \in \{ \ket{\uparrow}, \ket{\downarrow} \}$ to auxiliary states, is described by~\footnote{For simplicity we set $\hbar = 1$ in the supplemental material.}
\begin{align}
    \hat{H}_{\rm int} = \sum_i\Delta E_i \ket{i}\bra{i}
\end{align}
with~\cite{grimm_optical_2000}
\begin{align}
    \Delta E_i = \frac{\alpha_i(\lambda, \theta) I(x)}{2 c \epsilon_0}.
\end{align}
The polarizability $\alpha_i(\lambda, \theta)$ depends on the wavelength $\lambda$ and the polarization angle $\theta$ of the tweezer beam with respect to the magnetic field, $c$ is the speed of light and $\epsilon_0$ is the vacuum permittivity.
We define our two-level qubit subspace in the $^{40}$Ca$^+$ ion by $\ket{\downarrow} = \ket{S_{1/2}, m_j = +1/2}$ and $\ket{\uparrow} = \ket{D_{5/2}, m_j = +3/2}$, with respective polarizabilities
\begin{align}
    \frac{\alpha_\downarrow}{2 c \epsilon_0} &= -2\pi \times 4.81 \times 10^{-4}~\rm{Hz}/(\rm{W}/\rm{m}^2), \\
    \frac{\alpha_\uparrow}{2 c \epsilon_0} &= 2\pi \times 2.25 \times 10^{-4}~\rm{Hz}/(\rm{W}/\rm{m}^2),
\end{align}
given a linearly polarized light field along the $y$-axis. 
The calculations for these elements are performed using Ref.~\cite{atomphys}, combined with data from Ref.~\cite{kiruga_portal_2025}.
Note that the state-dependence in $\hat{H}_{\rm int}$ can also be written in terms of $\hat{\sigma}_z$, as used in Ref.~\cite{vasquez_state-dependent_2024}.

We approximate the optical tweezer field $I(x,y)$ as a Gaussian beam, where in the following we only consider the profile along the $x$-axis, which is given in the focal plane ($z=0$) as
\begin{align}
    I(x) = \frac{2P}{\pi w_0^2} \exp\left(-\frac{2x^2}{w_0^2}\right),
\end{align}
with $P$ the laser power and $w_0$ the beam waist.
We describe the motion of a single ion along the $x$-axis as a harmonic oscillator of frequency $\omega_x$ with the Hamiltonian
\begin{align}
    \hat{H}_m = \frac{\hat{p}^2}{2m} + \frac{1}{2}m\omega_x^2\hat{x}^2=  \omega_x \left( \hat{a}^\dagger \hat{a} + \frac{1}{2} \right),
\end{align}
where $\hat{a}^\dagger$ and $\hat{a}$ are the creation and annihilation operators and $m$ the mass of the ion.
The position and momentum operators are respectively given by
\begin{align}
    \hat{x} &= a_0 (\hat{a}^\dagger + \hat{a}), \\
    \hat{p} &= \frac{i}{2 a_0}(\hat{a}^\dagger - \hat{a}),
\end{align}
with $a_0 = \sqrt{1/(2m\omega_x)}$ the spatial extent of the motional
ground-state wavepacket.
The position of the ion can be written as $x = x_0 + \hat{x}$, where $x_0$ is the equilibrium position of the ion.
The total Hamiltonian of the system is then given by
\begin{align}
    \hat{H} = \hat{H}_m + \hat{H}_{\rm int}.
\end{align}
In order to derive the state-dependent Gaussian operators, we expand $\Delta E_i$ around $x_0$ up to the second order in $\hat{x}$, which gives
\begin{align}
    \Delta E_i 
    \approx \frac{\alpha_i}{2c\epsilon_0} 
    \bigg[ 
        I(x_0) + 
        \left. \partial_x I(x) \right|_{x_0} \hat{x} +
        \frac{1}{2} \left. \partial_x^2 I(x) \right|_{x_0} \hat{x}^2
    \bigg],
\end{align}
corresponding to the ac-Stark shift, force and curvature terms, which can be summarized as
\begin{align}
    \hat{H}_{\rm int} \approx \hat{H}_{\rm ac-Stark} + \hat{H}_{\rm force} + \hat{H}_{\rm curv}.
\end{align}
In our case the expansion up to second order is sufficient to describe the state-dependent Gaussian operations. 
Including higher order terms $\propto\hat{x}^3$ or $\propto\hat{x}^4$ leads to non-Gaussian operations, such as tri- or quad-squeezing operations and also a Kerr term, which can be neglected in our regime. 

\subsection{Motional mode shift}

The curvature term $\hat{H}_{\rm curv}$ leads to the state-dependent motional mode shift $\Delta \omega_{x, i}$, which can be derived using the relation~\cite{grimm_optical_2000}
\begin{align}
    \hat{H}_{\rm{curv}} = \frac{1}{2} m \omega_{\rm{tw}, i}^2 \hat{x}^2,
\end{align}
where $\omega_{\rm{tw}, i}$ denotes the optical trapping frequency of the ion provided by the tweezer.
We can explicitly determine
\begin{align}
    \omega_{\rm{tw}, i}^2 (x_0) = \frac{2\alpha_i(4x_0^2-w_0^2)}{c \epsilon_0 m w_0^4} I(x_0).
\end{align}
and use this expression to rewrite the Hamiltonian as
\begin{align}
    \hat{H}_m + \hat{H}_{\rm{curv}} = \frac{\hat{p}^2}{2m} + \frac{1}{2}m(\omega_x^2 + \omega_{\rm{tw}, i}^2)\hat{x}^2,
\end{align}
where we define $\omega_{x, i}'^{2} = \omega_x^2 + \omega_{\rm{tw}, i}^2$ and the observed motional mode shift as $\Delta \omega_{x, i} = \omega_{x, i}' - \omega_x$.
Assuming $\omega_{\rm{tw}, i}^2 \ll \omega_x^2$, we approximate
\begin{align}
    \Delta \omega_{x, i} \approx \frac{\omega_{\rm{tw}, i}^2}{2 \omega_x}.
\end{align}

\section{State-dependent Gaussian gate set operations}\label{sec:gaussian_gate_set}

\subsection{Tweezer intensity modulation}

To implement the state-dependent Gaussian operations we modulate the tweezer intensity at a frequency $\delta$,
\begin{align}
    I(x, t, \delta) &= \frac{I(x)}{2} \left(1 + \cos(\delta t)\right) \notag \\
    &= \frac{I(x)}{2} + \frac{I(x)}{4} \left(e^{i\delta t} + e^{-i\delta t}\right) \\
    &= I_{\rm stat}(x) + I_{\rm mod}(x, t, \delta). \notag
\end{align}
where we have decomposed the modulated intensity into a static contribution $I_{\rm stat}(x)$ and a time-dependent contribution $I_{\rm mod}(x, t, \delta)$.
The former produces a constant ac-Stark shift and motional mode shift and the latter resonantly drives the displacement or squeezing operation, depending on the choice of $\delta$.
The corresponding operators can be derived by transforming the total Hamiltonian into the interaction picture with respect to the time-independent contribution
\begin{align}
    \hat{H}_0 = \hat{H}_m + \hat{H}_{\rm int, stat},
\end{align}
where $\hat{H}_{\rm int, stat}$ denotes the static part of $\hat{H}_{\rm int}$.
In the regime $\omega_{\rm{tw}, i}^2 \ll \omega_x^2$, this yields the dressed motional frequencies
\begin{align}\label{eq:tweezer_approx}
    \omega_{x, i}' \approx \omega_x + \frac{\Delta \omega_{x, i}}{2},
\end{align}
where the factor $1/2$ reflects the time-averaged intensity $I_{\rm stat} = I(x)/2$ seen by the ion during modulation.
In this frame, the ladder operators evolve at the state-dependent dressed frequency,
\begin{align}
    \hat{a} &\rightarrow \hat{a} e^{-i \omega_{x, i}' t}, \\
    \hat{a}^\dagger &\rightarrow \hat{a}^\dagger e^{i \omega_{x, i}' t},
\end{align}
while $a_0$ is kept the same given the approximation~\ref{eq:tweezer_approx}.

\subsection{Displacement operation}

The displacement operation is implemented by modulating the tweezer intensity at the dressed motional frequency $\delta = \omega_{x, i}'$, which resonantly drives $\hat{H}_{\rm force}$,
\begin{align}
    \hat{H}_{\rm{force}}  &= \sum_i \frac{\alpha_i}{2c\epsilon_0} \left. \partial_x I_{\rm mod}(x, t, \omega_{x, i}') \right|_{x_0} \hat{x} \ket{i}\bra{i} \notag \\
    &= -\sum_i \frac{\alpha_i}{2c\epsilon_0} \frac{4x_0}{w_0^2} I_{\rm mod}(x_0, t, \omega_{x, i}') a_0 \notag \\
    &\quad \times (\hat{a} e^{-i \omega_{x, i}' t} + \hat{a}^\dagger e^{i \omega_{x, i}' t}) \ket{i}\bra{i}.
\end{align}
Applying the rotating wave approximation (RWA) and inserting into the unitary time evolution $\hat{U}_{\rm force}(t) = e^{-i \hat{H}_{\rm force} t}$ gives us the displacement operator
\begin{align}
    \hat{D}(\zeta_i) = \exp \left ( \zeta_i \hat{a}^\dagger - \zeta_i^* \hat{a} \right ).
\end{align}
The amplitude $\zeta_i$ can explicitly be determined as
\begin{align}
    \zeta_i = i\frac{\alpha_i a_0 x_0 I(x_0)}{2c\epsilon_0 w_0^2}t,
\end{align}
where $|\zeta_i|$ is maximal at $x_0= \pm w_0/2$ and zero at $x_0=0$.

\subsection{Squeezing operation}

Similar to the displacement operation, the squeezing operation is implemented by modulating the tweezer intensity at $\delta = 2 \omega_{x, i}'$, which resonantly drives $\hat{H}_{\rm curv}$.
Using $[\hat{a}, \hat{a}^\dagger] = 1$, we determine
\begin{align}
    \hat{H}_{\rm curv} &= \sum_i \frac{\alpha_i}{2c\epsilon_0} \frac{1}{2} \left. \partial_x^2 I_{\rm mod}(x, t, 2\omega_{x, i}') \right|_{x_0} \hat{x}^2 \ket{i}\bra{i} \notag \\
    &= \sum_i \frac{\alpha_i}{2c\epsilon_0} \frac{2 \left(4x_0^{2} - w_{0}^{2}\right)}{w_{0}^{4}} I_{\rm mod}(x_0, t, 2\omega_{x, i}') a_0^2 \notag \\
    &\quad \times \left[\hat{a}^2 e^{-2 i \omega_{x, i}' t} + \hat{a}^{\dagger 2} e^{2 i \omega_{x, i}' t} + 2 \hat{a}^\dagger \hat{a} + 1 \right ] \ket{i}\bra{i},           
\end{align}
which, after applying the RWA and inserting into the unitary time evolution, gives us the squeezing operator
\begin{align}
    \hat{S}(\xi_i) = \exp \left( \frac{1}{2}(\xi_i^{*}\hat{a}^{2} - \xi_i\hat{a}^{\dagger 2}) \right).
\end{align}
Here, $\xi_i = r_i e^{i \phi}$ with $\phi$ being the phase of the squeezing and $r_i$ the squeezing parameter.
We can explicitly determine
\begin{align}
    r_i = \frac{\alpha_i a_0^2 (4x_0^2-w_0^2)I(x_0)}{2 c\epsilon_0 w_0^4}t,
\end{align}
which is maximal at $x_0=0$ and zero at $x_0=\pm w_0/2$.

\subsection{Rotation operation}

The rotation operation does not require a modulation of the tweezer intensity $(\delta = 0)$ and is generated by $\hat{H}_{\rm curv}$, similar to the motional mode shift.
With the unmodulated tweezer intensity $I(x)$, the curvature term reads
\begin{align}
    \hat{H}_{\rm curv} &= \sum_i \frac{\alpha_i}{2c\epsilon_0} \frac{1}{2} \left. \partial_x^2 I(x) \right|_{x_0} \hat{x}^2 \ket{i}\bra{i} \notag \\
    &= \sum_i \frac{\alpha_i}{2c\epsilon_0} \frac{2 \left(4x_0^{2} - w_{0}^{2}\right)}{w_{0}^{4}} I(x_0) a_0^2 \notag \\
    &\quad \times \left[\hat{a}^2 e^{-2 i \omega_{x, i}' t} + \hat{a}^{\dagger 2} e^{2 i \omega_{x, i}' t} + 2 \hat{a}^\dagger \hat{a} + 1 \right ] \ket{i}\bra{i},
\end{align}
where, in contrast to the squeezing operation, the absence of a resonant drive leaves the $\hat{a}^2$ and $\hat{a}^{\dagger 2}$ terms fast-oscillating in the interaction picture, so that they average out under the RWA.
Keeping only the static term $\propto\hat{a}^\dagger \hat{a}$, the unitary time evolution gives us the rotation operator
\begin{align}
    \hat{R}(\varphi_i) = \exp \left ( -i\varphi_i\hat{a}^\dagger\hat{a}\right ).
\end{align}
The rotation angle $\varphi_i$ can be explicitly determined as
\begin{align}
    \varphi_i = \frac{2\alpha_i a_0^2 (4x_0^2-w_0^2)I(x_0)}{c\epsilon_0 w_0^4}t.
\end{align}

\subsection{Beamsplitter operation}

For the beamsplitter, we consider two motional modes, in our case the two axial modes (c.o.m\@.\ and str\@.)\ of a two ion crystal, which can be treated as two independent harmonic oscillators
\begin{align}
    \hat{H}_m = \omega_a \left ( \hat{a}^\dagger \hat{a} + \frac{1}{2} \right ) + \omega_b \left ( \hat{b}^\dagger\hat{b} + \frac{1}{2} \right ),
\end{align}
where $\hat{a}, \hat{a}^\dagger$ and $\hat{b}, \hat{b}^\dagger$ are the annihilation and creation operators of the respective modes, and $\omega_a$ and $\omega_b$ their respective frequencies.
We denote the corresponding position operators as $\hat{x}_a$ and $\hat{x}_b$ and rewrite $\hat{H}_{\rm curv}$ for the two ion case as
\begin{align}
    \hat{H}_{\rm curv} = \sum_{i, k} \frac{\alpha_i}{2c\epsilon_0} \frac{1}{2} \left. \partial_x^2 I_{\rm mod}(x, t) \right|_{x_{0, k}} \hat{x}_k^2 \ket{i}\bra{i},
\end{align}
with the ion $k \in \{1, 2\}$.

The normal coordinates $\hat{x}_a$ and $\hat{x}_b$ are related to the local coordinates $\hat{x}_1$ and $\hat{x}_2$ via~\cite{james_quantum_1998}
\begin{align}
    \hat{x}_1 &= \frac{1}{\sqrt{2}} (\hat{x}_a + \hat{x}_b), \\
    \hat{x}_2 &= \frac{1}{\sqrt{2}} (\hat{x}_a - \hat{x}_b),
\end{align}
which can be generalized to
\begin{align}
    \hat{x}_k = \frac{1}{\sqrt{2}} (\hat{x}_a + s_k \hat{x}_b),
\end{align}
with $s_1 = +1$ for ion 1 and $s_2 = -1$ for ion 2.
Explicitly calculating $\hat{x}_k^2$ gives us
\begin{align}
    \hat{x}_k^2 = \frac{1}{2} \left( \hat{x}_a^2 + \hat{x}_b^2 + 2 s_k \hat{x}_a \hat{x}_b \right),
\end{align}
leading to the cross term $\hat{x}_a \hat{x}_b$ required for the beamsplitter interaction.
The two terms $\hat{x}_a^2$ and $\hat{x}_b^2$ again result in state-dependent motional mode shifts during the beamsplitter interaction.
Similar to the single ion case, we move to the interaction picture with respect to the dressed motional frequencies $\omega_{a, i, k}'$ and $\omega_{b, i, k}'$ of the respective modes, for state $\ket{i}$ and ion $k$, transforming the ladder operators as
\begin{align}
    \hat{a} &\rightarrow \hat{a}\, e^{-i \omega_{a, i, k}' t}, \qquad
    \hat{a}^\dagger \rightarrow \hat{a}^\dagger e^{i \omega_{a, i, k}' t}, \\
    \hat{b} &\rightarrow \hat{b}\, e^{-i \omega_{b, i, k}' t}, \qquad
    \hat{b}^\dagger \rightarrow \hat{b}^\dagger e^{i \omega_{b, i, k}' t}.
\end{align}
This gives us the time-dependence of the cross term $\hat{x}_a \hat{x}_b$, written in terms of the ladder operators as
\begin{align}
    \hat{x}_a \hat{x}_b &= a_{0}b_{0} \bigg [
    \hat{a} \hat{b}\, e^{-i (\omega_{a, i, k}' + \omega_{b, i, k}') t}
    + \hat{a} \hat{b}^\dagger e^{-i (\omega_{a, i, k}' - \omega_{b, i, k}') t} \notag \\
    &\quad + \hat{a}^\dagger \hat{b}\, e^{i (\omega_{a, i, k}' - \omega_{b, i, k}') t}
    + \hat{a}^\dagger \hat{b}^\dagger e^{i (\omega_{a, i, k}' + \omega_{b, i, k}') t} \bigg ].
\end{align}
Modulating the tweezer intensity at the difference frequency $\delta = \omega_{a, i, k}' - \omega_{b, i, k}'$ and applying the RWA gives the effective beamsplitter operator
\begin{align}
    \hat{B}(g_{ik}) = \exp\left (\frac{1}{2} (g_{ik}\hat{a}^{\dagger}\hat{b} - g_{ik}^{*}\hat{a}\,\hat{b}^{\dagger} )\right ),
\end{align}
where the beamsplitter strength $g_{ik}$ can be determined as
\begin{align}
    g_{ik} = -i s_k \frac{\omega_{\rm{tw}, i, k}^{2}}{8 \sqrt{\omega_{a, i, k}' \omega_{b, i, k}'}}.
\end{align}
The $\pi$-time for a full swap between the two modes is given by $t_{\pi} = \pi / |g_{ik}|$.
Modulating the tweezer intensity at the sum frequency $\delta = \omega_{a, i, k}' + \omega_{b, i, k}'$ instead gives the two-mode squeezing interaction of equal strength.

\begin{figure*}
  \centering
  \includegraphics[]{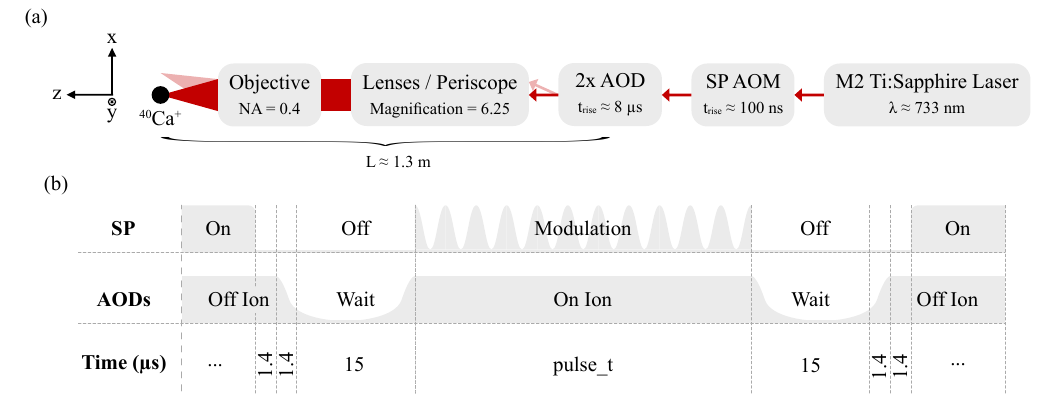}
  \caption{\justifying \textbf{Tweezer beam path and pulse sequence.} 
  (a) Beam path of the tweezer light (from right to left) starting at a Ti:sapphire laser, followed by a single-pass (SP) acousto-optical modulator (AOM) used for fast switching and amplitude modulation and delivered by an optical fiber to two crossed acousto-optical deflectors (AODs) used to position the tweezer beam onto the ion.
  After the AODs, the tweezer light is magnified by a set of lenses and directed onto the ion via a periscope, followed by a dichroic mirror and a $\rm{NA} = 0.4$ objective.
  The tweezer beam path measures $\approx \SI{1.3}{\meter}$ from the AODs to the ion.
  (b) Standard pulse sequence for a tweezer operation in order to minimize tweezer position drifts and to account for the slow rise time of the AODs.
  The ``Off Ion'' position of the AODs is $\approx \SI{60}{\micro\meter}$ displaced from the ``On Ion'' position.}
  \label{fig:tweezer_beam_path_and_pulse_sequence}
\end{figure*}

\section{Tweezer beam path and pulse sequence}\label{sec:tweezer_beam_path_and_pulse_sequence}

\subsection{Tweezer beam path}

Our experimental apparatus is the same as that of Ref.~\cite{leindecker_direct_2026}, with the main difference being that the optical tweezer light is generated by a Ti:sapphire laser tuned to $\lambda \approx \SI{733}{\nano\meter}$, with a single-pass (SP) acousto-optical modulator (AOM) used to control the tweezer light.
After the SP AOM, the tweezer light is steered relative to the ion using the same pair of crossed AODs as in Ref.~\cite{leindecker_direct_2026}.
The tweezer beam path, illustrated in Fig.~\ref{fig:tweezer_beam_path_and_pulse_sequence}~(a), is comparatively long, measuring $\approx \SI{1.3}{\meter}$ in free space from the AODs to the ion.
This likely degrades the beam-pointing stability, leading to the reduced motional coherence time we observe at the position of maximum tweezer gradient compared to the beam center.
Intensity modulation of the tweezer light is achieved by applying three rf tones to the SP AOM, producing a sinusoidal modulation of the intensity with a modulation index close to one.
The conversion between the AODs scan frequency and $\SI{}{\micro\meter}$ in the ion plane is found to be $\SI{0.160(1)}{\mega\hertz / \micro\meter}$.

\subsection{Pulse sequence}

Operating the tweezer at high powers of up to $\approx \SI{100}{\milli\watt}$ at the ion induces drifts in the positions of both the tweezer and the ion.
These drifts appear most strongly shortly after the tweezer is turned on and additionally depend on the pulse sequence.
We attribute their origin to charging and thermal effects along the tweezer beam path and in the ion trap.
To minimize the drifts, we keep the tweezer permanently on at maximum power, but displaced from the ion by $\approx \SI{60}{\micro\meter}$, as illustrated in the default pulse sequence in Fig.~\ref{fig:tweezer_beam_path_and_pulse_sequence}~(b).
We also perform regular position feedback of the tweezer with respect to the ion using low-power ac-Stark shift scans along the $x$- and $y$-axis, similar to Fig.~\ref{fig:position_dependence}~(a).

The rise time of the AODs of $\approx \SI{8}{\micro\second}$ is much longer than the rise time of the SP AOM of $\approx \SI{100}{\nano\second}$ (Fig.~\ref{fig:tweezer_beam_path_and_pulse_sequence}~(b)).
Therefore, we use the SP AOM for fast switching and amplitude modulation of the tweezer light and apply a wait time of $\SI{15}{\micro\second}$ when steering the AODs.

\begin{figure*}
    \centering
    \includegraphics[]{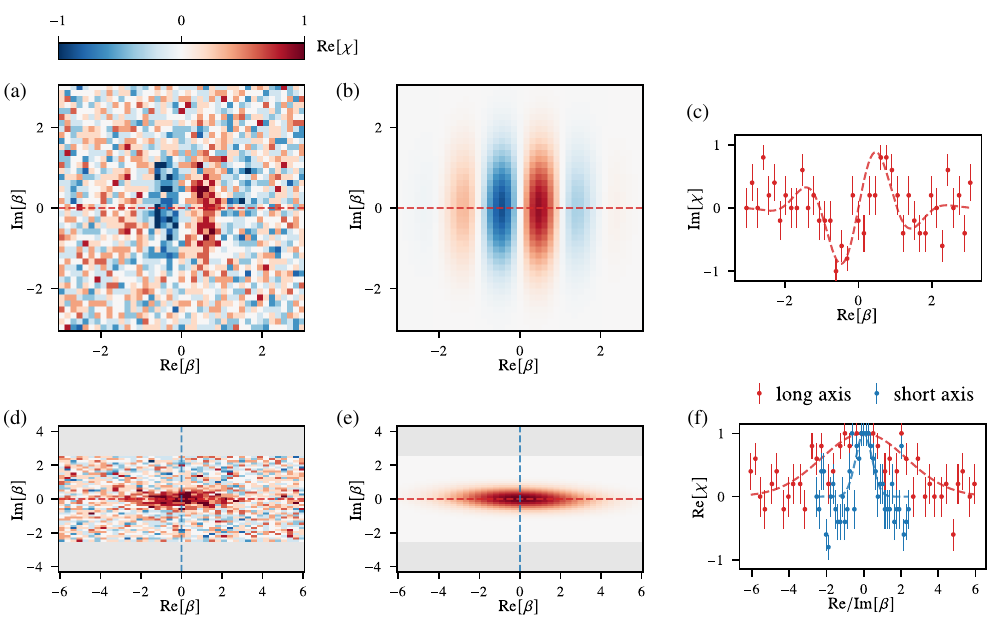}
    \caption{\textbf{Characteristic function fit.} 
    Raw data for displaced (a) and squeezed (d) states shown in Fig.~\ref{fig:state_dependence} for $\SI{20}{\micro\second}$ and $\SI{250}{\micro\second}$ operation time, respectively. 
    Subfigures (b) and (e) show the fit of the data used to extract the displacement and squeezing parameter. 
    For the displacement the slice perpendicular to the fringes (c) and for the squeezed state a slice along the short and long axes (f) are shown.}
    \label{fig:char_func_readout}
\end{figure*}

\section{Tomography of the motional states}\label{sec:tomography}

To determine the displacement amplitude $\zeta$, the squeezing strength $r$, and the coherent-state amplitude $\zeta_\mathrm{cat}$ of a cat state, we measure the characteristic function of the motion using a bichromatic SDF~\cite{fluhmann_direct_2020} and fit the corresponding ideal expression, adapted from~\cite{fluhmann_encoding_2019_phd_thesis}, to the data.

For a displaced state,
\begin{equation}
  \chi_D(\beta) = e^{-|\beta|^2/2}\,e^{\beta\zeta^* - \beta^*\zeta},
\end{equation}
with the complex displacement $\zeta$ as the fit parameter. 
For a squeezed state,
\begin{equation}
  \chi_S(\beta) = e^{-|\beta\cosh r + \beta^* e^{i\varphi}\sinh r|^2/2},
\end{equation}
the squeezing parameter $r$ and the squeezing angle $\varphi$ are treated as fit parameters. 
For a cat state,
\begin{equation}
  \chi_\mathrm{cat}(\beta) = \frac{1}{\mathcal{N}}
        \sum_{\substack{\delta,\varepsilon \\ \in\,\{\pm\zeta_\mathrm{cat}\}}}
        c_\delta^*\, c_\varepsilon\,
        e^{-\beta^*\varepsilon + \delta^*\beta + \delta^*\varepsilon
                    - \tfrac{|\delta|^2+|\beta|^2+|\varepsilon|^2}{2}},
\end{equation}
with normalization
\begin{equation}
  \mathcal{N} = \sum_{\substack{\delta,\varepsilon \\ \in\,\{\pm\zeta_\mathrm{cat}\}}}
    c_\delta^*\, c_\varepsilon\, e^{-\frac{|\delta|^2 + |\varepsilon|^2 - 2\delta^*\varepsilon}{2}},
\end{equation}
and coefficients $c_{\zeta_\mathrm{cat}} = 1$ and $c_{-\zeta_\mathrm{cat}} = e^{i\varphi_s}$, where $\varphi_s$ sets the relative phase between the two coherent components. 
Here the amplitude and phase of $\zeta_\mathrm{cat}$ together with $\varphi_s$ are fitted.

The characteristic function is reconstructed from the measured spin population via
\begin{equation}
    \mathrm{Re}(\chi_i) = 2\left(P_\downarrow - 0.5\right)
    \quad \text{or} \quad
    \mathrm{Im}(\chi_i) = 2\left(P_\downarrow - 0.5\right),
\end{equation}
where the real or imaginary part is selected by omitting or applying a $\pi/2$ pulse before the SDF~\cite{fluhmann_encoding_2019_phd_thesis}. 
The argument $\beta$ is set by the bichromatic pulse duration $c_\mathrm{sdf}$ through $|\beta| = 2c_\mathrm{sdf}t_\mathrm{sdf}$, with the SDF strength $t_\mathrm{sdf}$ treated as the fit parameter. 
Examples for a displaced and squeezed state are shown in Fig.~\ref{fig:char_func_readout}. 
The error of each fit was extracted from the covariance matrix of the fit.

\end{document}